\documentclass[12pt]{article}
\usepackage{epsfig}
 \newcommand{\beq}{\begin{equation}}
 \newcommand{\eeq}{\end{equation}}
 \newcommand{\beqa}{\begin{eqnarray}}
 \newcommand{\eeqa}{\end{eqnarray}}

 \newcommand{\mx}{\left[\begin{array}}
 \newcommand{\finmx}{\end{array}\right]}
 \def\vev#1{\langle #1 \rangle}
 \def\lsim{\raise0.3ex\hbox{$\;<$\kern-0.75em\raise-1.1ex\hbox{$\sim\;$}}}
 \def\gsim{\raise0.3ex\hbox{$\;>$\kern-0.75em\raise-1.1ex\hbox{$\sim\;$}}}
  
\long\def\ignore#1{}
 \evensidemargin=\oddsidemargin
\begin{document}       

 \begin{titlepage}
 \begin{flushright}
 hep-ph/0007266\\ CERN-TH/2000-218\\ IFIC/00-46\\ 
 UdeA-PE-00-04 \\ \today \\
 \end{flushright}
 \vspace*{5mm}
 \begin{center} 
   {\Large \bf Bilinear R-parity Violation and Small Neutrino Masses: a
     Self-consistent Framework
     }\\[15mm]
   {\large J. M. Mira${}^1$, E. Nardi${}^{1,2}$, D. A. Restrepo${}^3$,
     and J. W. F.  Valle${}^3$}\\
   \hspace{3cm}\\
   {\small ${}^1$Departamento de F\'\i sica, Universidad de Antioquia}\\
   {\small A.A 1226, Medellin, Colombia} \vskip.5cm {\small ${}^2$
     Theory Division, CERN, CH-1211 Geneva 23, Switzerland} \vskip.5cm
   {\small ${}^3$Instituto de F\'\i sica Corpuscular (IFIC), CSIC - U.
     de Val\`encia, \\ \small Edificio Institutos de Paterna, Apartado
     de Correos 22085\\ \small
     E-46071--Val\`encia, Spain}\\
   \hspace{3cm}\\
 \end{center}
 \vspace{5mm}
 \begin{abstract}
   We study extensions of supersymmetric models without R-parity which
   include an anomalous $U(1)_H$ horizontal symmetry.  Bilinear
   R-parity violating terms induce a neutrino mass at tree level
   $m_\nu^{\rm tree }\approx (\theta^2)^\delta\,$eV where
   $\theta\simeq 0.22$ is the $U(1)_H$ breaking parameter and
   $\delta$ is an integer number that depends on the horizontal
   charges of the leptons.  For $\delta=1$ a unique self-consistent
   model arises in which i) all the superpotential trilinear R-parity
   violating couplings are forbidden by holomorphy; ii) $m_\nu^{\rm
     tree }$ falls in the range suggested by the atmospheric neutrino
   problem; iii) radiative contributions to neutrino masses are
   strongly suppressed resulting in $\Delta m^2_{\rm solar}\approx$
   few $10^{-8}$ eV$^2$ which only allows for the LOW (or quasi-vacuum)
   solution to the solar neutrino problem; iv) the neutrino mixing
   angles are not suppressed by powers of $\theta$ and can naturally
   be large.
 \end{abstract}
 \end{titlepage}

 \section{Introduction}
 \label{sec:intro}

 The field content of the Standard Model (SM) together with the
 requirement of $G_{SM}=SU(2)_L\times U(1)_Y$ gauge invariance implies
 that the most general Lagrangian is characterized by additional
 accidental $U(1)$ symmetries implying Baryon ($B$) and Lepton flavor
 number ($L_i\,$, $i=e\,,\mu\,,\tau$) conservation at the
 renormalizable level.  When the SM is supersymmetrized, this nice
 feature is lost.  The introduction of the superpartners allows for
 several new Lorentz invariant couplings.  The most general
 renormalizable superpotential respecting the gauge symmetries reads
 \beq
 \label{WnoR}
 W=\mu_\alpha H_\alpha \phi_u + 
 \lambda_{\alpha\beta k} H_\alpha H_\beta l_k +
 \lambda'_{\alpha j k}    H_\alpha Q_j     d_k +
 \lambda''_{i j k}          u_i d_j d_k + 
 h^u_{jk}\phi_u Q_j u_k\,.  
 \eeq
 where $i,j,k=1\,,2\,,3$ and $\alpha,\beta=0,1\,,2\,,3\,$,
and all the fields appearing in (\ref{WnoR}) are superfields.
In the following we will denote with the same symbol the minimal
 supersymmetric SM (MSSM) superfields and their 
 SM fermion and scalar components.  Since we
 will soon extend the model to include a horizontal $U(1)_H$ symmetry,
 we take the fields in (\ref{WnoR}) in the basis where the horizontal
 charges are well defined.  We have denoted by $H_\alpha$ a vector
 containing the four hypercharge $Y=-1/2$ $SU(2)_L$ doublets of the
 MSSM and, without loss of generality,
 $H_0$ is the field whose main component is the down-type Higgs field:
 $H_0\simeq \phi_d\,$ ($\phi_d$ is defined as the direction in
 $H_\alpha$ field space that acquires a vacuum expectation).  It
 follows that $H_1\,,H_2$ and $H_3$ have as main components the lepton
 doublets $L_e\,,L_\mu$ and $L_\tau\,$, with $\langle L_i\rangle =0$
 by definition.  $\phi_u$ denotes the $Y=+1/2$ Higgs doublet,
 $u_i\,,d_j$ and $l_k$ ($i,j,k=1\,,2\,,3$) are the $SU(2)_L$ singlets
 up-type quarks, down-type quarks and leptons of the three
 generations, and $Q_j$ denotes the $SU(2)_L$ quark doublet.  The
 Yukawa couplings responsible of the up-type quark masses are denoted
 by $h^u_{jk}$ and, given our definition of the down-type Higgs field,
 in first approximation the leptons and down-type quarks Yukawa
 couplings are given by $ h^l_{jk}\simeq\lambda_{0jk}$ and
 $h^d_{jk}\simeq\lambda'_{0jk}\,$.  As it stands, (\ref{WnoR}) has
 potentially dangerous phenomenological consequences:
 \begin{itemize}
   
 \item[i)] The dimensionfull parameters $\mu_\alpha$ are gauge and
   supersymmetric invariant, and thus their natural value is expected
   to be much larger than the electroweak and supersymmetry breaking
   scales.  A large value of $\mu_0$ would result in too large
   Higgsino mixing term (this is the supersymmetric $\mu$ problem)
   while $\mu_i \sim \mu_0$ would give a large mass to one neutrino
   \cite{Hall:1984id,Ellis:1985gi,Banks:1995by}.
 
 \item[ii)] The dimensionless Yukawa couplings $h^l_{jk}\>(\simeq
   \lambda_{0jk})\,$, $h^d_{jk}\>(\simeq \lambda'_{0jk})$ and
   $h^u_{jk}$ are expected to be of order unity, suggesting that all
   the fermion masses should be close to the electroweak breaking
   scale.

 \item[iii)] The trilinear couplings
   $\lambda_{ijk}\,,\lambda'_{ijk}\,,\lambda''_{ijk}$ are also expected
   to be of order unity, implying unsuppressed $B$ and $L$ violating
   processes.

 \end{itemize}
 
 The approach originally suggested by Froggatt and Nielsen (FN)
 \cite{Froggatt:1979nt} to solve ii) and account for the fermion mass
 hierarchy turns out to be quite powerful in the context of the MSSM
 to solve also the $\mu $ problem.  FN postulated an horizontal
 $U(1)_H$ symmetry that forbids most of the fermion Yukawa couplings.
 The symmetry is spontaneously broken by the vacuum expectation value
 (vev) of a SM singlet field $\chi\,$ and a small parameter of the
 order of the Cabibbo angle $\theta =\langle\chi\rangle/M\simeq 0.22$
 (where $M$ is some large mass scale) is introduced.  The breaking of
 the symmetry induces a set of effective operators coupling the SM
 fermions to the electroweak Higgs fields, which involve enough powers
 of $\theta$ to ensure an overall vanishing horizontal charge.  Then
 the observed hierarchy of fermion masses results from the dimensional
 hierarchy among the various higher order operators.  When the FN idea
 is implemented within the MSSM, it is often assumed that the breaking
 of the horizontal symmetry is triggered by a single vev, for example
 the vev of the scalar component of a chiral supermultiplet $\chi$
 with horizontal charge $H(\chi)=-1\,$.  Then, because the
 superpotential is holomorphic all the operators carrying a negative
 charge are forbidden in the supersymmetric limit.  If under $U(1)_H$
 the bilinear term $H_0 \phi_u$ has a charge $n_0<0\,$, a $\mu_0$ term
 can only arise from the (non-holomorphic) K\"ahler potential,
 suppressed with respect the supersymmetry breaking scale $m_{3/2}$
 as~\cite{Giudice:1988yz}
 \beq\label{mu0}
 \mu_0\simeq m_{3/2}\>\theta^{|n_0|}\,.
 \eeq
 A too large suppression ($|n_0| > 1$) would result in unacceptably
 light Higgsinos, so that in practice on phenomenological grounds $n_0
 = -1$ is by far the preferred value.

 More recently it has been realized that the FN mechanism can play a
 crucial role also in keeping under control the trilinear $B$ and $L$
 violating terms in (\ref{WnoR}) without the need of introducing an ad
 hoc R-parity quantum
 number~\cite{Mira:2000fx,Binetruy:1996xk,Chun:1996xv,others,%
   Choi:1997se,Joshipura:2000sn}.  For example in~\cite{Mira:2000fx} it
 was argued that under a set of mild phenomenological assumptions about
 the size of neutrino mixings a non-anomalous $U(1)_H$ symmetry
 together with the holomorphy conditions implies the vanishing of all
 the superpotential $B$ and $L$ violating couplings.  A systematic
 analysis on the restrictions on trilinear R-parity violating couplings
 in the framework of $U(1)_H$ horizontal symmetries was also recently
 presented in~\cite{Joshipura:2000sn}.
 
 In this paper we argue that if the $\mu_0$ problem is solved by the
 horizontal symmetry in the way outlined above, and if the additional
 bilinear terms $\mu_i$ are also generated from the K\"ahler potential
 and satisfy the requirement of inducing a neutrino mass below the eV
 scale, as indicated by data on atmospheric
 neutrinos~\cite{Fukuda:1998mi,Fornengo:2000sr}, then in the basis
 where the horizontal charges are well defined, all the trilinear
 R-parity violating couplings are automatically absent.  This hints at
 a self-consistent theoretical framework in which R-parity is violated
 only by bilinear terms that induce a tree level neutrino mass in the
 range suggested by the atmospheric neutrino anomaly, $L$ and $B$
 violating processes are strongly suppressed, and the radiative
 contributions to neutrino masses are safely small so that $m_\nu^{\rm
   loop}\approx 10^{-4}\,$eV, which barely allows for the LOW or
 quasi-vacuum solutions to the solar neutrino
 problem~\cite{qva,updateGonzalez-Garcia:2000aj}.

 \section{Tree level neutrino mass}
 \label{sec:treelevel}

 Our theoretical framework is defined by the following assumptions: \ 
 i) Supersymmetry and the gauge group $G_{SM}\times U(1)_H$.  \ ii)
 $U(1)_H$ is broken only by the vev of a field $\chi$ with horizontal
 charge $-1$; the field $\chi$ is a SM singlet, chiral under $U(1)_H$.
 \ iii) The ratio between the vev $\vev{\chi}$ and the mass scale $M$
 of the FN fields is of the order of the Cabibbo angle $\theta \simeq
 \vev{\chi}/M \simeq 0.22$.  In the following we will denote a field and
 its horizontal charge with the same symbol, e.g.  $H(l_i)= l_i$ for
 the lepton singlets, $H(Q_i)= Q_i$ for the quark doublets, etc.  It is
 also useful to introduce the notation $f_{ij} =f_i-f_j$ to denote the
 difference between the charges of two fields.  For example $H_{i0}$
 denotes the difference between the charges of the $H_i\simeq L_i$
 `lepton doublet' and the $H_0\simeq\phi_d$ `Higgs field'.  On
 phenomenological grounds we will assume that the charge of the $\mu_0$
 term is $n_0 = -1\,$  and we will also assume negative 
 charges $n_i=H_i+\phi_u < n_0$
 for the other three bilinear terms $H_i\phi_u\,$.  It is worth
 stressing that the theoretical constraints from the cancellation of
 the mixed $G_{SM}\times U(1)_H$ anomalies hint at the same value
 $n_0=-1$ both in the anomalous~\cite{Nir:1995bu} and in the
 non-anomalous \cite{Mira:2000fx} $U(1)_H$ models (see section
 \ref{Theory}).  With the previous assumptions the four components of
 the vector $\mu_\alpha$ in (\ref{WnoR}) read
 \beq\label{mu}
 \mu_\alpha \simeq m_{3/2}\,
 (\theta^{|n_0|},\theta^{|n_1|},\theta^{|n_2|},\theta^{|n_3|})\,,
 \eeq
 where coefficients of order unity multiplying each entry have been
 left understood.  It is well known that if $\mu_\alpha$ and the vector
 of the hypercharge $Y=-1/2$ vevs $v_\alpha \equiv \vev{H_\alpha}$ are
 not aligned \cite{Hall:1984id,Banks:1995by}:
 \beq \label{sinxi}
 \sin\xi \equiv \frac{\mu\wedge v}{\sqrt{
 v_\alpha v^\alpha\>\mu_\beta \mu^\beta \ }}\neq 0
 \eeq
 the neutrinos mix with the neutralinos \cite{ProjectiveMassMatrix},  
 and one neutrino mass is
 induced at the tree level \cite{Banks:1995by}:
 \begin{equation}\label{mnu}
   m_\nu^{\rm tree} \simeq \frac{\mu
     \cos^2\!\beta\,}{\sin2\beta\,\cos\xi-\frac{\mu
       M_1M_2}{M_Z^2M_\gamma}}\,\sin^2\!\xi \,,
 \end{equation}
 where $M_\gamma=M_1\cos^2\theta_W+M_2\sin^2\theta_W\,$, $M_1$ and
 $M_2$ are the $U(1)_Y$ and $SU(2)_L$ gaugino masses, and $\tan\beta =
 \vev{\phi_u}/\vev{\phi_d}\,$.  
 Since $m_b/\vev{\phi_u} \tan\beta \approx \theta^{2.7}
 \tan\beta$ (with $m_b(m_t)\sim 2.9\,$GeV \cite{Fusaoka:1998vc}) in
 the following we will use the parameterization
 $\tan\beta=\theta^{\>x-3}$ that ranges between 90 and 1 for $x$
 between 0 and 3.  Keeping in mind that we are always neglecting
 coefficients of order unity, we can approximate $\cos^2\beta
 =(1+\tan^2\beta)^{-1}\approx \theta^{\>2\, (3-x)}$. Taking also
 $M_1\simeq M_\gamma\,$, $\mu M_2 / M_Z^2\gg \sin2\beta\,\cos\xi\,$
 and $100\,$GeV$\lsim M_2\lsim$ $500\,$GeV we obtain from (\ref{mnu})
 \begin{equation}\label{sinxinum}
   m_\nu^{\rm tree} 
 \approx \left[\theta^{\,-(5+x)}\> \sin\xi\right]^2 \,{\rm eV}\,.
 \end{equation}
The magnitude of the tree-level neutrino mass as a function of
$\log_\theta\sin\xi \approx H_{30}$ for different 
values of $x$ (which in our notations parameterizes $\tan\beta$) 
is illustrated in fig. 1.  The
grey bands correspond to equation (\ref{mnu}) with $M_2$ ranging
between $100\,$GeV and $500\,$GeV, while the dashed lines correspond
to the approximate expression (\ref{sinxinum}).
In general, two conditions have to be satisfied to ensure exact
$\mu_\alpha$--$v_\alpha$ alignment and $m_\nu^{\rm tree}=0\,$
\cite{Banks:1995by}: 1) $\mu_\alpha \propto B_\alpha\,$ and 2)
$\tilde m^2_{\alpha \beta}\mu_\beta = \tilde m^2 \mu_\alpha\,$, where
$B_\alpha$ is the bilinear soft-breaking term coupling the $H_\alpha$
and $\phi_u$ scalar components, and $\tilde m^2_{\alpha \beta}$ is the
matrix of the soft scalar masses for the $H_\alpha$ fields.

In our case the goodness of the alignment between $\mu_\alpha$ and
$v_\alpha$ is controlled by the horizontal symmetry, and in particular
there is no need of assuming universality of the soft breaking terms
to suppress $m_\nu^{\rm tree}$ to an acceptable level.  This is
because the previous two conditions are automatically satisfied in an
approximate way up to corrections of the order $\theta^{|H_{i0}|}\,$,
where the minimum charge difference between $H_0$ and the $H_i$
`lepton' fields is responsible for the leading effects.  Thus we can
estimate
\beq\label{sinxith}
\sin\xi \approx  
\theta^{|H_{i0}|} =
\theta^{|n_i-n_0|} 
\simeq \frac{\mu_i}{\mu_0}\,.
\eeq
Confronting (\ref{sinxith}) with (\ref{sinxinum}) it follows that in
order to ensure that $m_\nu^{\rm tree}$ is {\it parametrically}
suppressed below the eV scale we need
\beq \label{condition}
|n_i-n_0| > 5+x \qquad (i=1,2,3)\,.
\eeq
\begin{figure}
\centerline{\protect\hbox{
\psfig{file=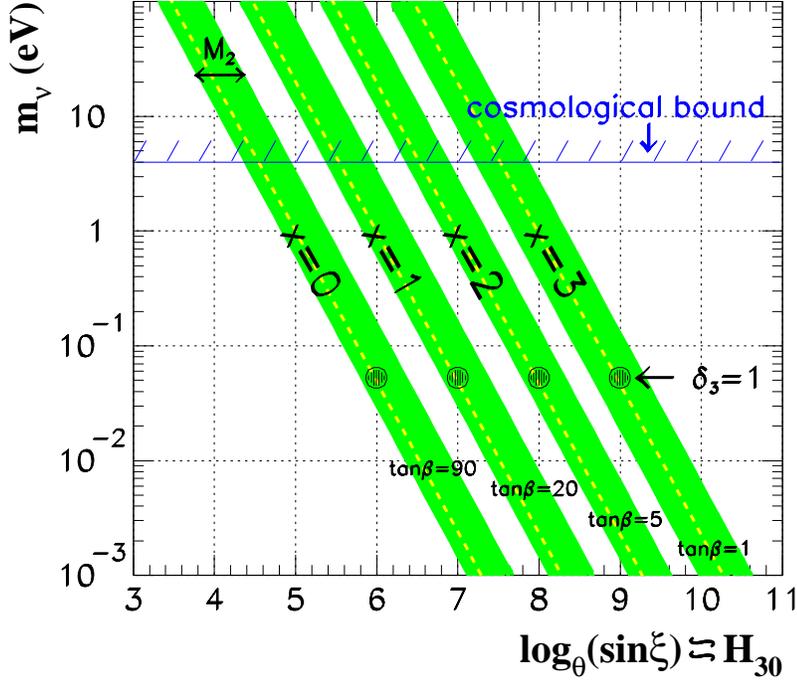,height=10cm}}}
\caption[]{Tree-level neutrino mass dependence 
on $\log_\theta\sin\xi  \approx H_{30}$ for different 
assignments of the charge difference $H_{30}=H_3-H_0$ and 
for different values of $\tan\beta$.  Details in  the text.}
\end{figure}

\section{Vanishing of the $\lambda$ and $\lambda'$
  couplings}

As we have shown in the previous section, requiring a sufficient
suppression of tree level neutrino mass with respect to the Higgsino
mass implies that the charges $H_i$ should be much larger in absolute
value than $H_0$.  Then it follows that in the basis where the charges
are well defined, the relations $H_0\simeq \phi_d$ and
$h^{l\,(d)}_{ij}\simeq \lambda^{(\prime)}_{0ij}$ are satisfied to a
very good approximation.  Let us introduce the parameterization
\beq \label{condition2}
|n_i-n_0| - (5+x)= \delta_{i} \,.
\eeq
Without loss of generality, we can also assume $n_1\leq n_2\leq n_3$
which implies
\beq\label{del23}
m_\nu^{\rm tree} \approx  \theta^{\,2\delta_3}\>{\rm eV}\,. 
\eeq
It is worth stressing that the parameter that controls the scaling of
$m_\nu^{\rm tree}$ with respect to changes in the values of the
horizontal charges is $\theta^{\,2}\simeq 0.05\,$, and thus neutrino
masses are much more sensitive to the horizontal symmetry than the
other fermion masses that scale with $\theta$.  For example
$\delta_3=-1$ yields $m_\nu^{\rm tree} \sim 20\,$eV in conflict
with cosmological structure formation \cite{Silk}; $\delta_3=0$ yields
$m_\nu^{\rm tree}\sim 1\,$eV which implies a sizeable amount of hot
dark matter; however, as we will see, it also allows for non-vanishing
$\lambda$ and $\lambda''$ couplings; for $\delta_3=1$ all the
trilinear R-parity violating couplings are forbidden, and at the same
time $m_\nu^{\rm tree}\sim 5\times 10^{-2}\,$ eV (see Fig.~1) is in
the correct range for a solution to the atmospheric neutrino
problem~\cite{Fukuda:1998mi,Fornengo:2000sr}; finally, $\delta_3=2$
would suppress $m_\nu^{\rm tree}$ too much to allow for such a
solution. 


Let us now write the down-quarks and lepton Yukawa matrices as
\beqa
\label{Yukawa}
h^d_{jk}&\simeq&\theta^{H_0+Q_j+d_k}=\theta^{Q_{j3}+d_{k3}+x}\,,
\qquad \nonumber   \\
h^l_{jk}&\simeq&\theta^{H_0+H_j+l_k}=\theta^{H_{j3}+l_{k3}+x}\,,
\eeqa
where $x=H_0+Q_3+d_3=H_0+H_3+l_3\,$ consistently with our
parameterization of $\tan\beta$ and 
with the approximate equality between
the bottom and tau masses at sufficiently high energies (which in
particular allows for $b$--$\tau$ 
Yukawa unification).  The order of
magnitude of the trilinear R-parity violating couplings is then:
\beqa
\label{Trilinear}
\lambda'_{ijk}&\simeq&\theta^{n_i-n_0}\,h^d_{jk} \simeq
\theta^{Q_{j3}+d_{k3}-(5+\delta_i)}\,, \nonumber \\
\lambda_{ijk}&\simeq&\theta^{n_i-n_0}\,h^l_{jk} \simeq 
\theta^{H_{j3}+l_{k3}-(5+\delta_i)}\,.
\eeqa
One can show that the phenomenological information on the charged
fermion mass ratios and quark mixing angles \beqa \label{hierarchy}
m_u:m_c:m_t &\simeq& \theta^{\,8}:\theta^{\,4}:1\,, \nonumber \\
m_d:m_s:m_b &\simeq&
\theta^{\,4}:\theta^{\,2}:1\,,  \nonumber \\
m_e:m_\mu:m_\tau &\simeq&
\theta^{\,5}:\theta^{\,2} :1\,,   \nonumber \\
V_{us}\simeq \theta\,, & \quad & V_{cb}\simeq \theta^{\,2}\,, \eeqa
which gives rise to eight conditions on the fermion charges\footnote{
  Note that $V_{ub} \simeq V_{us}V_{cb}\simeq \theta^{\,3}$ is a
  prediction of the model (in agreement with the experimental
  measurements) and does not give additional constraints.}  can be
re-expressed in terms of the following sets of eight charge
differences
\cite{Binetruy:1996xk,Choi:1997se,Leurer:1993wg,Binetruy:1995ru,Dudas:1995yu}

 \begin{center}
\beq
\bigskip
\begin{tabular}{|c|cccccc|c|cc|}
 \hline
\rm model  & $Q_{13}$ & $Q_{23}$ & $d_{13}$ & $d_{23}$ & $u_{13}$ & $u_{23}$
&\rm model & $H_{13}+l_{13}$ & $H_{23}+l_{23}$  \\ 
 \hline
\strut MQ1: \ &\  3  & 2  &  1 & 0 &  5 & 2 & \ ML1: \ & 5 &\  2 \\
\strut MQ2: \ & --3  & 2  &  7 & 0 & 11 & 2 & \ ML2: \ & 9 & --2 \\
\hline
\end{tabular}
\label{Models}
\bigskip
\eeq
 \end{center}
 We will not repeat here the phenomenological analysis leading to
 these sets of charge differences, since this has been extensively
 discussed in the literature \cite{Binetruy:1996xk,Choi:1997se,
 Leurer:1993wg,Binetruy:1995ru,Dudas:1995yu}; however, let us comment
 briefly on the different models listed in (\ref{Models}).  The first
 set of charge differences labeled as MQ1 and ML1 corresponds the
 simplest solution where all the charges are fixed before
 supersymmetry breaking by the phenomenological conditions listed in 
 (\ref{hierarchy}).
%
 Note however, that the charge differences in the second row labeled
 as MQ2 and ML2 are also compatible with (\ref{hierarchy}).  This is
 due to the fact that in MQ2 and ML2 some entries in the mass matrices
 have negative values of the charges, and initially correspond to
 holomorphic zeroes.
%
 After canonical diagonalization of the field kinetic terms these
 zeroes are lifted to non-vanishing values which are the correct ones
 to reproduce the same pattern (\ref{hierarchy}) of mass ratios and
 quark mixing angles \cite{Binetruy:1996xk,Choi:1997se,Dudas:1995yu}.
%
 For example, with the restriction $x\neq 3\,$, the overall charge of
 the $(1,2)$ entry in the down quark mass matrix
 $Q_{13}+d_{23}+x=-3+x$ is negative and implies
 $h^d_{12}=0\,$. However, after $Q_i$ and $d_j$ field redefinition
 this entry is lifted to to $h^d_{12}\simeq \theta^{\, x+3}$
 \cite{Binetruy:1996xk} which yields the correct value of the Cabibbo
 mixing angle $V_{us}\simeq \theta\,$.  Similarly, with the
 restriction $x\neq2,3$ ML2 reproduces correctly the lepton mass
 ratios in (\ref{hierarchy}).

 Confronting now (\ref{Trilinear}) with (\ref{Models}) we can conclude
 the following

\begin{itemize}
  
\item In MQ1, $\delta_i\geq 0\,$ is a sufficient condition to ensure
  that the overall charges of the $\lambda'$ couplings are negative,
  implying that in the charge basis all these couplings are forbidden
  by holomorphy.
  
\item 
  In ML1, $\delta_i \geq 1$ is only a necessary condition to achieve
  $\lambda_{ijk}=0\,$.  Since in the leptonic sector the single values
  of the charge differences that control the mixing angles are not
  known, we need more assumptions to make a definite statement about
  these couplings.  Let us note that the values $H_{12}=-1,-2$ are
  always excluded since they would result in incorrect values of the
  $m_e/m_\mu$ mass ratio, while $H_{12}=-3$ is allowed only for $x=0$.
  Therefore in the leptonic sector the condition $n_i<0$ forces the
  mixing between the first two generation neutrinos $V_{e\mu}\simeq
  \theta^{|H_{12}|}$ to be either very strongly suppressed ($\lsim
  \theta^3\,$) or of order unity.  The first case excludes the
  possibility of explaining the solar neutrino data through
  $\nu_e$-$\nu_\mu$ oscillations. The other possibility $H_{12}=0$
  corresponds to $\nu_e$-$\nu_\mu$ mixing not suppressed by powers of
  $\theta\,$, and hence gives the possibility of implementing a large
  mixing angle solution for the solar neutrino problem.  On the other
  hand, since a maximal $\nu_\mu$--$\nu_\tau$ mixing is strongly
  supported by the atmospheric neutrino data, we will take $H_{23}=0\,$
  as a phenomenological assumption. Then from eq. (\ref{Trilinear}) it
  is easy to see that $H_{23}=H_{12}=0$ is enough to guarantee the
  vanishing of all the $\lambda_{ijk}$ couplings.
  
\item In MQ2, $Q_{23}+d_{13}=9\,$ so that to eliminate the $\lambda'$
  couplings we would need $\delta_i \geq 5\,$.  This results in a very
  large suppression of the tree level neutrino mass $m_\nu^{\rm tree}
  \lsim 10^{-7}\,$eV so that this case is not very interesting from
  the point of view of neutrino phenomenology.  Insisting on
  $\delta_i=1\,$ results in $\lambda'_{i21}\simeq \theta^{\,3} $ and
  $\lambda'_{i31}\simeq \theta \,$ while all the others $\lambda'$
  couplings vanish.  Apparently, this is not in conflict with the
  existing experimental limits.  However, after $Q_i$ and $d_i$ field
  redefinition a tiny coupling $\lambda'_{i12}\simeq 
  \lambda'_{i31}\theta^{\,|Q_{13}|+|d_{12}|} \simeq \theta^{\,11}$ is
  generated. This is enough to conflict with the strong
  limit $\lambda'_{i21}\lambda'_{i12}\lsim \theta^{\,15}\,$ from
  $K$--$\bar K$ mixing \cite{Barbieri:1986ty}.  We conclude that in
  MQ2 either the neutrino masses are uninterestingly small, or the
  $\lambda'$ conflicts with existing experimental limits~\footnote{As
  we will see in the next section, MQ2 with $\delta_i=1$ is also
  excluded by the requirement that the $\lambda''$ couplings vanish.}.
  
\item In ML2, once we set $H_{23}=0$ to allow for maximal
  $\nu_\mu$--$\nu_\tau$ mixing, the
  lepton mass ratios (\ref{hierarchy}) can be correctly reproduced
  only if $H_{12}\geq 4\,$, which would again exclude the possibility
  of explaining the solar neutrinos deficit through 
  $\nu_e$--$\nu_\mu$ oscillations.

\end{itemize}

In conclusion, we have shown that in the framework of models of
Abelian horizontal symmetries, the phenomenological information on the
charged fermion mass ratios and quark mixing angles listed in
(\ref{hierarchy}) and re-expressed in terms of the eight horizontal
charge differences in (\ref{Models}), when complemented with the
requirement that $m_\nu^{\rm tree}$ is adequately suppressed below the
eV scale ($\delta_i \geq 1$) hints at one self-consistent model
(MQ1+ML1) where all the $\lambda$ and $\lambda'$ couplings vanish.  It
is interesting to note that $\delta_3=1$ which yields $m_\nu^{\rm
  tree}\approx \theta^2$~eV in the correct range required by the
atmospheric neutrino problem is also the minimum value that ensures
$\lambda=0\,$, $\lambda'=0$ and, as we will see in the next section,
$\lambda''=0\,$.

\section{Vanishing of the $\lambda''$ couplings}

Even if the trilinear lepton number violating couplings are absent in
the basis where the horizontal charges are well defined, field
rotation to the physical basis $(\phi_d,L_i)$ will still induce tiny
$\delta \lambda$ and $\delta \lambda'$ terms.  In general the
couplings induced in this way remain safely small to satisfy most of
the experimental constraints, however some combination of the
$\delta\lambda'$ with the B violating $\lambda''$ couplings can
endanger proton stability.  In this section we will show that the
additional theoretical constraints from cancellation of the mixed
$G_{SM}\times U(1)_H$ anomalies, which are mandatory if $U(1)_H$ is a
local symmetry, ensure that all the $\lambda''$ charges are negative
and that the couplings are forbidden by holomorphy.\footnote{Here we
  assume that the $U(1)_H$ is anomalous, so that the anomaly
  cancellation is achieved via the Green-Schwarz
  mechanism~\cite{Green:1984sg}. This is the only possibility
  consistent with the implicit assumption $m_u\neq 0$ made in
  (\ref{hierarchy}) \cite{Mira:2000fx}. A study of the non-anomalous
  case is presented in \cite{Usnew}.}  Since for the $\lambda''$ a
change of basis or a field redefinition cannot lift any of the
holomorphic zeroes, proton stability is not in jeopardy.

Let us introduce the notation $n_Q = \sum_{i} Q_i $ for the sum of the
charges of the quark doublets and let us write the charge of a generic
$\lambda''_{ijk}$ coupling as
\beq\label{lambdapp}
d_i+d_j+u_k = d_{i1}+d_{j2}+u_{k3}+
(Q_1+d_1+H_0)+
(Q_2+d_2+H_0)+
\phi_u -n_Q-2n_0\,, 
\eeq
where we have used $Q_3+u_3+\phi_u=0$ as implied by $m_t\sim 
\vev{\phi_u}$.  The consistency conditions for cancellation of the
anomalies via the Green-Schwarz mechanism \cite{Green:1984sg} imply
that the coefficients of the mixed $SU(2)_L^2\times U(1)_H$ and
$SU(3)_C^2\times U(1)_H$ anomalies $C_2=\sum_\alpha H_\alpha+\phi_u+3
n_Q $ and $C_3 = \sum_i\,(2Q_i+d_i+u_i) $ must be equal
\cite{Ibanez:1993fy}. This equality can be written as
\beq \label{C2C3}
\sum_{\alpha=0}^3 n_\alpha+3(n_Q-\phi_u) =  3(6+x-n_0)
\eeq
where for $C_2$ on the left-hand side of (\ref{C2C3}) we have used
$\sum_{\alpha} H_\alpha= \sum_{\alpha} n_\alpha-4\phi_u\,$, and the
expression for $C_3$ on the right-hand side can be easily derived from
the charge differences given in (\ref{Models}) and holds for both MQ1
and MQ2.

Inserting in (\ref{lambdapp}) the value of $\phi_u - n_Q$ derived from
the anomaly cancellation condition (\ref{C2C3}) and writing the
explicit values of the $m_d\,$ and $m_s\,$ charges appearing inside
the parentheses in (\ref{lambdapp}) (respectively $4+x$ and $2+x$ ) we
obtain
\beq\label{lambdapp2}
d_i+d_j+u_k = d_{i1}+d_{j2}+u_{k3}+
(x-n_0)+\frac{1}{3}\sum_{\alpha=0}^3 n_\alpha
\leq  d_{i1}+d_{j2}+u_{k3} - 5 - \frac{1}{3}\,,
\eeq
where in the last step we have used $n_0=-1$ and $n_1\leq n_2\leq
n_3\leq -(6+x)$ as suggested by the analysis in the previous sections.
Now it is straightforward to verify that the charge differences in
(\ref{Models}) imply $d_{i1}+d_{j2}\leq 0\,$ both in MQ1 and MQ2
(remember that $i\neq j$ because of the antisymmetry of the
$\lambda''$) and $u_{k3}\leq 5$ (MQ1), $u_{k3}\leq 11 $ (MQ2).  The
values that saturate these relations are the most conservative ones.
Therefore in MQ1 $d_i+d_j+u_k<0\,$ for all values of the indices and
independently of $\tan\beta$, thus ensuring the vanishing of all the
$\lambda''$ couplings, while in MQ2 some of the $\lambda''$ 
do not vanish.

\bigskip

\section{One loop neutrino masses}
\label{sec:oneloop}

It has long been realized that loop effects may lead to radiative
neutrino masses~\cite{Ross:1985yg}. In order to estimate the 
size of these effects in the present framework, first we need to 
evaluate the $\delta\lambda$ and $\delta\lambda'$ terms induced by
the rotation from the basis $(H_0,H_i)$ in which the charges are well
defined to the basis $(\phi_d,L_i)$ in which the Yukawa couplings are
well defined.
Given that $H_0\simeq \phi_d + \sum_i \theta^{\,|H_{i0}|} L_i$ we obtain
\beqa
\label{induced}
(\delta\lambda')_{ijk}&\simeq& \theta^{\,|H_{i0}|}\, h^d_{jk}
          \simeq 
\theta^{\,5+\delta_i+x}\, \theta^{\,Q_{j3}+d_{k3}+x} \,,\\
(\delta\lambda)_{ijk}&\simeq& \theta^{\,|H_{i0}|}\, h^l_{jk}
          \simeq 
\theta^{\,5+\delta_i+x}\, \theta^{\,H_{j3}+l_{k3}+x} \,.
\eeqa
Once non-vanishing $\lambda$ and $\lambda'$ couplings are generated,
quark-squark and lepton-slepton loop diagrams will induce a mass for
the two neutrinos that are massless at the tree
level~\cite{Hempfling:1996wj,Borzumati:1996hd,Nardi:1997iy,Romao:2000up}.
An approximate expression for the leading one-loop contributions to the
neutrino mass matrix reads~\cite{Dimopoulos:1988jw}
\begin{equation} \label{mnuloop}
(m^{\rm loop}_{\nu})_{ij} \simeq  
 {3\,  (\delta\lambda^\prime)_{i k l}
(\delta\lambda^\prime)_{j m n} \over 8 \pi^2}\,
{(m^d)_{k n}(\tilde  M^{d^{\scriptstyle \,2}}_{LR})_{l m}\over\tilde  m^2}\,
+ \, { (\delta\lambda)_{i k l} (\delta \lambda)_{j m n} 
\over 8 \pi^2}\,
{(m^l)_{k n}
(\tilde  M^{l^{\scriptstyle \,2}}_{LR})_{l m}\over\tilde  m^2}\,.
\end{equation}
Here $m^d$ ($m^l$) is the $d$--quark (lepton) mass matrix, $\tilde
M^{d(l)^{\scriptstyle \,2}}_{LR}$ is the left--right sector in the
mass-squared matrix for the $\tilde d$ ($\tilde l$) scalars, $\tilde
m$ represents a slepton or squark mass, and the expression holds at
leading order in $\tilde M^2_{LR} / \tilde m^2\,$.  As was discussed
in \cite{Borzumati:1996hd} the largest loop contribution comes from
quark-squark loops involving $(m^d)_{32}\sim (m^d)_{33}\sim m_b$ and $
({\tilde M}^{d^{\scriptstyle \,2}}_{LR})_{32} \sim ({\tilde
  M}^{d^{\scriptstyle \,2}}_{LR})_{33} \sim \tilde m\, m_b\,$, and
gives a mass of the order
\beq
(m_{\nu}^{\rm loop})_{ij}
\approx 
\frac{3}{8\pi^2}\,\frac{m^2_b}{\tilde m}\,
 (\delta\lambda^\prime)_{i 33}\, (\delta\lambda^\prime)_{j 33}\>  
\approx \> 
\theta^{\delta_i+\delta_j+4x}\> {{\rm eV}}\,,  
\eeq
where we have used $3/(8\pi^2)\,(m_b/\tilde m)\,(m_b/1\,$eV) $\approx
\theta^{-10}\,$ corresponding to $\tilde m \approx 100\,$GeV.  We see
that for $\delta_2=\delta_3=1$ (that allows for a
$\nu_\mu$--$\nu_\tau$ mixing angle without parametric suppression) we
have two main possibilities: {\it (i)} $x=0$ ($\tan\beta\sim m_t/m_b$)
and $m_\nu^{\rm loop}\approx m_\nu^{\rm tree}\approx \theta^2 \sim \,$ few
$10^{-2}$ eV.  While this allows for a $m^2_{\nu_\tau}-m^2_{\nu_\mu}$
difference in the correct range for the atmospheric neutrino problem,
$\nu_e$--$\nu_\mu$ oscillations do not solve the solar neutrino
problem.  Only for $\tilde m \gsim 1\,$TeV we obtain enough
suppression and $m_\nu^{\rm loop}\sim \> $few~$10^{-3}\,$eV can fall in
the correct range for the large mixing angle solutions to the solar
neutrino problem. Of course, $x=0$ implies that the value of
$\tan\beta$ is very large ($\gsim 60$) and therefore this case is
phenomenologically disfavored \cite{Grossman:1994ax,Barger:1993ac}.
{\it (ii)} $x=1$ ($\tan\beta \approx 10$-$40\,$) yields $m_\nu^{\rm
  loop} \approx \theta^6 \sim 10^{-4}\>{\rm eV}$ which besides fitting
the atmospheric neutrino mass squared difference, also allows for the
LOW or the quasi-vacuum solution to the solar neutrino problem.
Finally $x=2$ ($\tan\beta\sim 5$) would yield a too large suppression
$m_\nu^{\rm loop} \approx \theta^{10} \sim 10^{-7}\>{\rm eV}$ to be
interesting for the solar neutrinos.

In conclusion, our analysis results in the following set of 
fields charge differences and of $n_\alpha=H_\alpha+\phi_u$ 
charge sums:

 \begin{center}
\beq
\label{qn}
\bigskip
\begin{tabular}{|cccccc|cccc|cc|}
 \hline
 $Q_{13}$ & $Q_{23}$ & $d_{13}$ & $d_{23}$ & $u_{13}$ & $u_{23}$
           & $H_{13}$ & $H_{23}$ & $l_{13}$ & $l_{23}$ & $n_i   $ & $n_0 $\\ 
 \hline
\strut  3  & 2  &  1 & 0 &  5 & 2  
        & 0  & 0  &  5 & 2 & $-8$& $-1$ \\
\hline
\end{tabular}
\label{Model}
\bigskip
\eeq
 \end{center}

\noindent
where we have used the value $x=1$ 
(corresponding to $\tan\beta \approx 10$--$40\,$) as 
suggested by the analysis of the loop effects. The corresponding
structure of the charged fermion mass matrices is: 
\begin{equation}
  \label{eq:udmatrices} 
  \frac{M^u}{\langle\phi_u\rangle}\simeq\left[
    \begin{array}{ccc}
      \theta^{8}&\theta^{5} &\theta^{3}\\
      \theta^{7}&\theta^{4} &\theta^{2}\\
      \theta^{5}&\theta^{2} &1
    \end{array}\right],\ \ 
  \frac{M^d}{\langle\phi_d\rangle}\simeq\theta\left[\begin{array}{ccc}
      \theta^{4}&\theta^{3} &\theta^{3}\\
      \theta^{3}&\theta^{2} &\theta^{2}\\
      \theta&1&1
    \end{array}\right],\nonumber \ \ 
  \frac{M^l}{\langle\phi_d\rangle}\simeq\theta
  \left[
    \begin{array}{ccc}
      \theta^{5}&\theta^{2} &1\\
      \theta^{5}&\theta^{2} &1\\
      \theta^{5}&\theta^{2} &1
    \end{array}
  \right]. 
\end{equation} 
%
%
In the Appendix we will derive the individual charges of an anomaly 
free model that reproduces these results.

\section{Inputs versus Predictions}
\label{Theory}

Models based on a single $U(1)_H$ Abelian factor are completely
specified in terms of the horizontal charges of the SM fields.  There
are five charges for each fermion family plus two charges for the
Higgs doublets, for a total of 17 charges that {\it a priori} can be
considered as free parameters (the charge of the $U(1)_H$ breaking
parameter $\theta$ is just a normalization factor).  The individual
value of these charges is determined by a set of
phenomenological and theoretical conditions. To some extent it is a
matter of taste what is taken as an input condition, and what is
derived as a model prediction.  However it is important to understand
to what extent the model has a predictive power, and to what extent it
just has enough freedom to fit the experimental data. The purpose of
this section is to clarify this issue.

The six mass ratios plus two CKM mixing angles listed in
(\ref{hierarchy}) provide the first eight constraints on the fermion
charges.  There are two additional constraints from the absolute
values of the masses of the third generation fermions, corresponding
to a top mass unsuppressed with respect to the electroweak scale and
to the approximate equality between the $b$ and $\tau$ masses at
high energy 
\beqa\label{absolute}
m_t \sim \vev{\phi_u}  &\Longrightarrow& Q_3+u_3+\phi_u =0 \\
m_b \sim m_\tau\  &\Longrightarrow& 
x\equiv Q_3+d_3+H_0=H_3+l_3+H_0 \nonumber \,. 
\eeqa
In this paper we have also assumed that the supersymmetric $\mu$ 
problem is solved by the horizontal symmetry and we have taken 
the phenomenologically preferred value of the charge 
of the $\mu$ term 
\beq\label{n0}
n_0 = H_0+\phi_u = -1
\eeq
as an additional input.  If we assume that $U(1)_H$ is a gauge
symmetry, then additional constraints arise from the requirement of
cancellation of the mixed $G_{SM}\times U(1)_H$ anomalies.  The
vanishing of the coefficient of the $U(1)_Y\times U(1)_H^2$ anomaly
quadratic in the horizontal charges
\beq\label{quadratic}
C^{(2)}= \phi_u^2  -\sum_\alpha H^2_\alpha +
\sum_i \left[Q_i^2- 2 u_i^2 + d_i^2 +\ell_i^2\,\right]
\eeq
gives a first condition.  If, as we are assuming here, the
non-vanishing mixed anomalies linear in the horizontal charges are
canceled through the Green-Schwarz mechanism by a $U(1)_H$ gauge shift
of an axion field $\eta(x)\to \eta(x) -\xi(x)\,\delta_{GS}$
\cite{Green:1984sg} the following consistency condition must be also
satisfied \cite{Ibanez:1993fy}
\beq
\label{ibanez}
{C_3}={C_2}=\frac{C_1}{k_1}=\delta_{GS},
\eeq
where $C_1=\phi_u+\sum_\alpha H_\alpha+\frac{1}{3}
\sum_i[Q_i+8u_i+2d_i-3l_i] $ is the coefficient of the mixed
$U(1)_Y^2\times U(1)_H$ anomaly, and $C_2$ and $C_3$ have been defined
before eq.(\ref{C2C3}).  While the first equality in (\ref{ibanez})
represents an additional constrain on the horizontal charges, the
second condition depends on the hypercharge normalization factor $k_1$
that, since we are not postulating any GUT symmetry, must be
considered as a new arbitrary parameter.  When written explicitly in
terms of horizontal charges, eq. (\ref{ibanez}) yields the following
interesting relation \cite{Mira:2000fx}:
\beq
\label{GScondition}
n_0+\eta_l-\eta_d = (k_1 -\frac{5}{3})\> \delta_{GS}/2\,, 
\eeq
where we have introduced the notation $\eta_d\equiv \sum_i
(Q_i+d_i+H_0)\simeq \log_\theta(\det M^d/\vev{\phi_d})$ and $\eta_l$ is
defined in a similar way.  From the fermion mass ratios in
(\ref{hierarchy}) we obtain $\eta_l-\eta_d=1\,$ that, together with
the assumption (\ref{n0}) implies $k_1=5/3\,$.  Therefore, while the
second equality in (\ref{GScondition}) does not provide additional
constraints on the horizontal charges, it predicts gauge coupling
unification for the canonical value $\sin^2\theta_W=3/8$.  Of course,
we could have equivalently taken the running of the gauge couplings in
the MSSM as a good reason to assume canonical gauge couplings
unification \cite{Nir:1995bu}, then $n_0=-1$ would have resulted as a
theoretical prediction.  In summary, the 17 horizontal charges are
constrained by eleven phenomenological conditions (including $n_0=-1$)
and by two theoretical conditions from anomaly cancellation.  This
leaves us with four free parameters, and we can chose them to be the
charges $n_i$ ($i=1,2,3$) of the bilinear terms $\mu_i$, and
$x=Q_3+d_3+H_0$ that fixes the value of $\tan\beta\,$.  The
expressions of the horizontal charges for all the SM fields 
as a function of these four parameters is given in the Appendix.

The main predictions of the model is the vanishing of all the trilinear
R-parity violating couplings in the charge basis, as well as $x=1$
that corresponds to $\tan\beta$ in the range $\approx10$--$40\,$.
In what concerns the pattern of neutrino mixings, our model is most
naturally realized with no parametric suppression of the mixing
angles, in agreement with the solar and atmospheric neutrino
observations, and in sharp contrast with the pattern of mixings in the
quark sector.  The exact values of the mixings depend on the unknown
coefficients of order unity, which are not determined by the Abelian
symmetry.  Finally, the absence of parametric suppression also applies
to the mixing angle which is restricted by reactor neutrino
experiments~\cite{Apollonio:1999ae}, whose small value 
in the present framework can only arise from a conspiracy between 
the unknown coefficients of order unity.

\section{Conclusions}

We have studied extensions of supersymmetric models without R-parity
which include an anomalous horizontal symmetry.  We have assumed that
all the bilinear superpotential terms coupling the up-type Higgs
doublet with the four hypercharge $-1/2$ doublets carry negative
horizontal charges, and hence are forbidden by holomorphy.  We have
constrained the value of these charges by several theoretical and
phenomenological requirements, such as having an acceptable Higgsino
mass ($\mu$ problem) and neutrino masses suppressed below the
electron-volt scale, as suggested by present neutrino data.  We have
found that under these conditions all the trilinear R-parity violating
superpotential couplings vanish, yielding a consistent model which is
defined by the charge differences in (\ref{qn}), where lepton number
is mildly violated only by small bilinear terms.  The model allows for
neutrino masses in the correct ranges suggested by the atmospheric
neutrino problem and by the LOW and quasi-vacuum solutions to the
solar neutrino problem.  However, no precise theoretical information
can be obtained about the neutrino mixing angles except for the fact
that, unlike the quark mixings, there is no parametric suppression of
their values and thus they can be naturally large.


\section*{Acknowledgements}

We thank J. Ferrandis for discussions. 
E. N. acknowledges the IFIC-CSIC of the University 
of Val\`encia where most of this work was carried out 
for the pleasant hospitality and for financial support. 
This work was supported by
DGICYT grant PB98-0693 and by the EEC under the TMR contract
ERBFMRX-CT96-0090. J.M.M and D.A.R are supported by COLCIENCIAS

\appendix
\renewcommand{\theequation}{\thesection.\arabic{equation}}
\def\sectionapp#1{\setcounter{equation}{0}
\section{#1}}

\sectionapp{Appendix} 
\label{sec:appendixa}

In this Appendix we derive the general expressions for 
the individual field charges satisfying the set of   
13 phenomenological and theoretical constraints 
corresponding to     
the six mass ratios for the quarks and the charged leptons  
plus the two quark mixing angles listed in (\ref{hierarchy});  
the two relations provided by 
the absolute value of the masses of the third generation 
fermions given in (\ref{absolute});   
one phenomenological assumption about the charge of 
the $\mu$ term (\ref{n0});  
one theoretical constraint corresponding to the consistency 
conditions (\ref{ibanez}) for the coefficients of 
the mixed linear anomalies (the second constraint fixes $k_1=5/3$) 
and one additional  constraint from the vanishing  
of the mixed anomaly quadratic in the horizontal charges (\ref{quadratic}).
As discussed in section 6, this leaves us with 
four free parameters that we choose to be
$n_i$ ($i=1,2,3$) and $x$.  
We obtain 
\begin{eqnarray}
  Q_3&=& \frac1{15\,\left( 7 + x  \right) }
  \Big[-180 - 45x - 3x^2 + Q_{13}(41 + 5x)
  -7L_{23}  + L_{23}^2
  \nonumber\\
  &&\hspace{1.5cm} 
  + n_1( 2 + x + L_{23} )
  + n_2( 9 + x - L_{23} )  
  + n_3(9 + x) 
  \Big],  \label{eq:q3} \\  
  H_3&=&\frac1{15
    \left( 7 + x  \right) }
  \Big[20  + 50x + 6x^2+ 18Q_{13} - 21 L_{23} + 3 L_{23}^2
  \nonumber\\
  &&\hspace{1.5cm}
  - n_1(  29 + 2x - 3 L_{23}) 
  -n_2( 8 + 2x +3 L_{23}) 
  +n_3(97+13x)
  \Big],   \label{eq:l3} \\\nonumber
\end{eqnarray}
where $L_{23}=H_{23}+l_{23}$ and 
$Q_{13}$ parametrize the two different possibilities  
for the quark and lepton charge differences given 
in (\ref{Models}).
In terms of $Q_3$ and $H_3$ and of our  
four free parameters we have 
\begin{equation}
  \begin{array}{lll}
    \phi_u&=&n_3-H_3\\
    H_0&=&-1+\phi_u
  \end{array}
  \qquad\qquad\qquad
  \begin{array}{lll}
    u_3&=&-Q_3-\phi_u\\
    d_3&=&-Q_3+x-H_0\\
    l_3&=&-H_3+x-H_0
  \end{array}
\end{equation}
and from these all the other individual charges can be 
straightforwardly determined from
the charge differences in eq.~(\ref{Models}).
The solution for the charges in model MQ1+ML1 
for the preferred values $n_1=n_2=n_3=-8$ and $x=1$ 
is given  in Table~\ref{tab:1}.
\begin{table}[ht]
  \begin{center}
    \begin{footnotesize}
      \begin{tabular}{|ccccccccccccccccc|}\hline
        &&&&&&&&&&&&&&&&\\
        $Q_1$&$Q_2$&$Q_3$&$u_1$&$u_2$&$u_3$&$d_1$&$d_2$&$d_3$&$H_1$&
        $H_2$&$H_3$&$l_1$&$l_2$&$l_3$&$\phi_u$&$H_0$\\ 
        $\frac{161}{30}$&$\frac{131}{30}$&$\frac{71}{30}$
        &$\frac{103}{15}$&$\frac{58}{15}$&$\frac{28}{15}$
        &$-\frac{18}{5}$&$-\frac{23}{5}$&$-\frac{23}{5}$
        &$-\frac{113}{30}$&$-\frac{113}{30}$&$-\frac{113}{30}$
        &$\frac{98}{15}$&$\frac{53}{15}$&$\frac{23}{15}$
        &$-\frac{127}{30}$&$\frac{97}{30}$\\
         &&&&&&&&&&&&&&&&\\\hline
      \end{tabular}
    \end{footnotesize}
    \caption{The anomaly free set of charges 
 of  model MQ1+ML1 for $x=1$ and  $n_1=n_2=n_3=-8$} 

    \label{tab:1}
  \end{center}
\end{table}

\end{document}